\documentclass[runningheads,a4paper]{llncs}

\usepackage{amssymb}
\usepackage{graphicx}

\usepackage{tikz,pgfplots}
\pgfplotsset{compat=1.14}

\usepackage{url}
\urldef{\mailsa}\path{brendon.boldt1@marist.edu}

\newcommand{\tabhorzspacing}{0.5em}
\newcommand{\tabvertspacing}{1.1}

\begin{document}

\mainmatter  % start of an individual contribution

\title{Using LSTMs to Model the Java Programming Language\footnotetext{This is a pre-print of an article published in Artificial Neural Networks and Machine Learning -- ICANN 2017. The final authenticated version is available online at: \url{https://doi.org/10.1007/978-3-319-68612-7_31}}}

\titlerunning{Using LSTMs to Model the Java Programming Language}

\author{Brendon Boldt}
\authorrunning{Brendon Boldt}

\institute{Marist College,\\
3399 North Rd. Poughkeepsie, NY, USA\\
\mailsa\\
}


\toctitle{Lecture Notes in Computer Science}
\tocauthor{Authors' Instructions}
\maketitle

\begin{abstract} 
Recurrent neural networks (RNNs), specifically long-short term memory 
networks (LSTMs), can model natural language effectively. This research 
investigates the ability for these same LSTMs to perform next ``word'' 
prediction on the Java programming 
language. Java source code from four different repositories
undergoes a transformation that preserves the logical structure of the 
source code and removes the code's various specificities such as 
variable names and literal values. Such datasets and an additional English 
language corpus are used to train and test standard LSTMs' ability to predict
the next element in a sequence. 
Results suggest that LSTMs can effectively model
Java code achieving perplexities under $22$ and accuracies above $0.47$, which
is an improvement over LSTM's performance on the English language which demonstrated
a perplexity of $85$ and an accuracy of $0.27$. This research can have
applicability in other areas such as syntactic template suggestion and automated bug
patching.
\end{abstract} 

\section{Introduction}
\label{submission}

Machine learning techniques of language modeling are often applied
to natural languages, but techniques used to model natural languages
such as $n$-gram, graphed-based, and context sensitive models
can be applicable to programming languages as well
\cite{Allamanis} \cite{Nguyen} \cite{Asaduzzaman}.
One such application of a language model is next-word prediction
which can prove very useful for tasks such as syntactic template
suggestion and bug patching
\cite{Nguyen} \cite{Kim}.
There has been research into programming language models
which use Bayesian statistical inference ($n$-gram models)
to perform next-word prediction \cite{Allamanis}.
Yet some of the most successful natural language models have been
built using recurrent neural networks (RNNs); their ability to
remember information over a sequence of tokens makes them particularly apt for
next-word prediction \cite{Zaremba}.

Specifically, long-short term memory (LSTM) RNNs have further improved
the basic RNN model by increasing the ability of an RNN to remember
data over a long sequence of input without the signal decaying
quickly \cite{Zaremba}. LSTMs are a sequence-to-word language
model which means given a sequence of words (e.g., words in the
beginning of a sentence), the model will produce a probability
distribution describing what the next word in the sequence is.

In terms of the Java programming language,
we are specifically investigating next-statement prediction in method
bodies. While other parts of Java source code (e.g., class fields,
import statements) do have semantic significance, method bodies make up
the functional aspect of source code\footnote{
Functional insofar as method bodies describe the active (non-declarative)
behavior of the program.} and most resemble natural language sentences.
Just as individual semantic tokens (words) comprise natural language
sentences, statements, which can be thought of as semantic tokens,
comprise method bodies. Furthermore, the semantics of individual natural
language words coalesce to form the semantics of sentence just as the
semantics of the statement in a method body form the semantics of the
method as a whole. By this analogy, language modeling techniques which
operate on sentences comprised of words could apply similarly to method
bodies comprised of statements.

%~%~% Also \subsubsection{}

%should be given.  See Section~\ref{final author} for details of how to

\section{Tokenizing Java Source Code}

We are specifically looking at predicting the syntactic structure  of the next 
statement in within Java source code method bodies. The syntactic structure 
of a complete piece of source code can be represented as an abstract 
syntax tree (AST) where each node of the tree represents a distinct 
syntactic element (e.g., statement, boolean operator, literal integer). 
Method bodies are, in particular, comprised of statements which, more or 
less, represent a self-contained action. Each of these statements is the 
root of its own sub-AST which represents the syntactic structure of only 
that statement. In this way, statements are 
independent, semantically meaningful units of a method body which are suitable 
to be tokenized for input into the RNN.

Nguyen et al. \cite{Nguyen} studied a model for syntactic statement prediction 
called ASTLan which uses Bayesian statistical inference to interpret and 
predict statements in the form of sequential statement ASTs. While Bayesian 
statistical inference can be applied to statements directly in their AST 
form, RNNs operate on independent tokens such as English words. Thus, it is 
necessary that statement ASTs be flattened into a tokenized form in order to 
produce an RNN-based model.

\subsection{Statement-Level AST Tokenization}

The RNN model described in Zaremba et al. \cite{Zaremba} specifically uses space-delimited 
text strings; hence, when the statement ASTs are tokenized, they 
must be represented as space-delimited text strings.

To show the tokenization of Java source, take the following Java statement:
\texttt{int x = obj.getInt();}.
The corresponding AST, as given by the Eclipse AST parser, appears in Figure \ref{ast-figure} \cite{Eclipse}.
This statement, in turn, would be transformed as follows\footnote{
\texttt{VariableDeclarationStatement} is not included in the
tokenized version of the AST since the syntax is adequately represented
by starting with the root node's children.}:

\begin{figure}
\begin{center}
\centerline{\includegraphics[height=25mm]{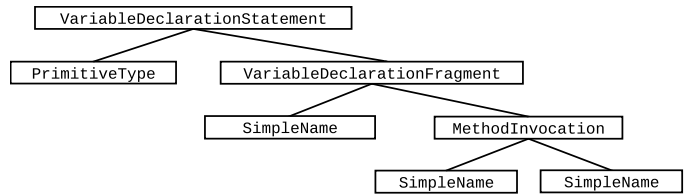}}
\caption{The abstract syntax tree (AST) representation of of the Java statement
\texttt{int x = obj.getInt();}}
\label{ast-figure}
\end{center}
\vskip -7mm
\end{figure}

\begin{verbatim}
    _PrimitiveType_VariableDeclarationFragment(_SimpleName
    _MethodInvocation(_SimpleName_SimpleName))) 
    
    _60(_39_59(_42_32(_42_42)))
\end{verbatim}

The first token uses the AST node names while second token
represents the same AST by instead using
integer IDs corresponding to the AST node names as assigned by
the Eclipse parser
(e.g., $60$ corresponds to ``PrimitiveType'' nodes and $42$ corresponds to
``SimpleName'' nodes). Using integer IDs saves space and is the format
used in the actual LSTM.

Individual AST nodes
are separated by underscores (``\texttt{\_}'') and parentheses are used
to denote a parent-child relationship so that the tree structure of
the statement is preserved. In fact, it is possible  to recreate the
syntax of the original source  code from the tokens; thus, this 
tokenization is lossless in terms of  \textit{syntactical} information 
yet lossy in other areas. For example, variable and function names are
discarded during the translation to make the model independent of
variable and function names.

\subsection{Method-Level Tokenization}

Consider the following Java method:

\begin{verbatim}
    int foo() {
        int x = obj.getInt();
        if (x > 0) {
            x = x + 5;
        }
        return x;
    }
\end{verbatim}

Each statement in the method body is tokenized just as the single statement 
was above, and the resulting tokens are delimited using spaces. Braces, while not 
statements, are included (denoted by ``\texttt\{'' and
``\texttt\}'') to retain the semantic structure of the method body. 
%It is the case that the return type and parameters are included as the first 
%token with a leading ``\texttt('' to denote that it is a method
%signature.
The method above becomes the following sequence of tokens:

\begin{verbatim}
    (_39_42 { _60(_39_59(_42_32(_42_42)))
     _25(_27(_42_34) { _21(_7(_42_27(_42_
     34))) } _41(_42) } 
\end{verbatim}

The sequence of these tokens forms a ``sentence'' which  
the represents body of a Java method.
Sentences in the dataset are separated by the \texttt{<eos>} metatoken
to mark the end of a sentence.
These sentences of
tokens will then comprise the corpus that the LSTM network uses to train 
and make predictions.

\subsection{English and Java Source Corpora Used}

Similarly to Zaremba et al. \cite{Zaremba}, we are using the Penn Treebank (PTB) for the
English language corpus as it provides an effective, general sample of the English
language.
For the Java programming languages, four different corpora were each built
by processing (as described above) a large repository of Java source code. The
repositories used were
the Java Development Kit (JDK), Google Guava, ElasticSearch, and Spring Framework.
The JDK is a good reference for Java since it is a widely-used implementation
of the Java language; the other three projects were selected based on their
high popularity on GitHub in addition to the fact they are
Java-based projects.

It is important to note that the
PTB does not contain any punctuation while the tokenized
Java source contains ``punctuation'' only in the form of statement
body-delimiting curly braces (``\texttt\{'' and ``\texttt\}'')
since these are integral to the semantic structure of source code.
All English and Java corpora use a metatoken to mark the end of a sentence.

\subsection{Vocabulary Comparison}

In addition to preserving the logical structure of the source code,
another goal of the specific method of tokenization was to
produce a vocabulary with a frequency distribution similar to that of
the English corpus. If the same
Java statement tokens appear too frequently, the tokenization might be
generalizing the Java source too much such that it loses the underlying
semantics. If the statement tokens, instead, all have a very low frequency
it would be difficult to effectively perform inference on the sequence of 
tokens within the allotted vocabulary size.

\begin{figure}
\centering
\begin{tikzpicture}
	\begin{axis}[
	    height=45mm,
	    width=8cm,
		xlabel=Word Rank,
		ylabel=Frequency,
		xmin=0,
		xmax=30]
		\addlegendentry{PTB}
		\addplot [dashed, black] table [
	    	x=rank,
		    y=PTB,
		    mark=none,
		    col sep=tab] {freq.dat};
		\addlegendentry{JDK}
		\addplot [black] table [
	    	x=rank,
		    y=JDK,
		    mark=none,
		    col sep=tab] {freq.dat};
	\end{axis}
\end{tikzpicture} 
\caption{Comparison of English and Java word frequency distributions.
    The $y$-axis represents the total proportion of the word with a given
    rank (specified by the $x$-axis).}
\label{english-frequency}
\end{figure}
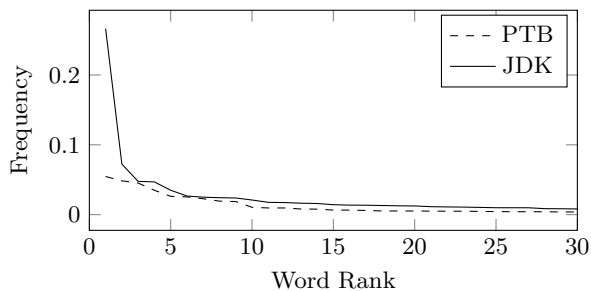

In all of the Java corpora, the left and right curly braces comprise
approximately $35\%$
of the total tokens present. This a disproportionately high number in
comparison to the rest of the tokens, but removing them from the frequency
distribution, since they classify as punctuation, gives a more accurate
representation of the vocabularies. The adjusted frequency distribution
shown in Figure \ref{english-frequency} compares the PTB to the
JDK source code. The rate of occurrence for the highest
ranked words is significantly higher in the JDK than in the PTB, but
the frequency distributions track closely together beyond the fifth-ranked
words.
Generally, all four Java corpora showed similar frequency distributions.

The statistical similarities between the English and the translated Java
corpora suggest that the Java statement tokens have an adequate amount of
detail in terms of mimicking English words. If the Java statement tokens
were too detailed, their frequencies would be far lower than those of English
words; if the Java statement tokens were not detailed enough, their
frequencies would be much higher than those of English words.

\begin{table}[t]
    \caption{Proportion and rank of the metatoken 
    \texttt{<unk>}. Proportions and ranks are from the adjusted Java corpora
    with the left and right curly braces removed.}
    \label{sample-table}
    \begin{center}
    \setlength{\tabcolsep}{\tabhorzspacing}
    {\renewcommand{\arraystretch}{\tabvertspacing}
    \begin{tabular}{lcc}
    \hline
    Corpus & Proportion & Rank\\
    \hline
    PTB                 & $0.0484$ & $2$ \\
    JDK                 & $0.0724$ & $2$ \\
    Guava               & $0.0476$ & $5$ \\
    ElasticSearch       & $0.1618$ & $2$ \\
    Spring Framework    & $0.0873$ & $2$  \\
    \hline
    \end{tabular}}
    \end{center}
    \vskip -7mm
\end{table}

Another consideration when comparing the English and Java corpora is the
prevalence of the metatoken \texttt{<unk>} which denotes a token not contained
in the language model's vocabulary.
Due to the nature of LSTMs, the vocabulary of the language model is finite;
hence, any word not contained in the vocabulary is considered unknown.
We specifically used a vocabulary size of $10,000$. A vocabulary size which
is too small will fail to represent enough words in the corpus; the result
is the LSTM seeing a high proportion of the \texttt{<unk>} metatoken. A
vocabulary which is too large increases the computation required during
training and inference.
The proportion of \texttt{<unk>} tokens in both the English and the Java
source data sets (save for ElasticSearch\footnote{ElasticSearch had a
proportion of $16\%$}) are $<10\%$ which indicates that
the $10,000$ word vocabulary accounts for approximately $90\%$ of the corpus'
words by volume. It is important that the Java corpora's \texttt{<unk>}
proportion is not significantly higher than that of the PTB since
that would suggest that $10,000$ is too small a vocabulary size to describe
the tokenized Java source code.

\section{Language Modeling}
\label{language-modeling}

\subsection{Neural Network Structure and Configuration}

In order to make a good comparison between language modeling in English
and Java, a model with demonstrated success at modeling English was
chosen. The model selected was an LSTM neural
network, a type of RNN, as described in
Zaremba et al. \cite{Zaremba}. This particular LSTM uses regularization via 
dropout to act as a good language model for natural languages
such as English \cite{Zaremba}.

The LSTM's specific configuration was the same as the ``medium''
configuration described in Zaremba et al. \cite{Zaremba} with the exception
that the data was trained for $15$ epochs instead of $39$ epochs.
Beyond $15$ epochs (on both the English and Java datasets), the 
training cost metric (perplexity) continued to decrease while the
validation cost metric remained steady. This suggests that the model
was beginning to overfit the training data and that further training
would not improve performance on the test data.
Specifically, this model contains two RNN layers with a vocabulary
size of $10,000$ words.

Each corpus was split into partitions such that $80\%$ was training data
and the remaining $20\%$ was split evenly between test and validation
data. Perplexity, the performance metric of the LSTM, is determined by the
ability of the LSTM to perform sequence-to-word prediction on the test
set of that corpus. Perplexity represents how well the prediction (in the
form of a probability distribution) given by the LSTM matches the actual
word which comes next in the sentence. A low perplexity means that the
language model's predicted probability distribution matched closely the
actual probability distribution, that is, it was better able to predict
the next word. Perplexity is the same metric that is used in
Zaremba et al. \cite{Zaremba} to compare language models.

\subsection{Language Model Metrics}

We chose word-level perplexity was chosen as the metric for comparing the
language models' performance on the given corpora since it provides
a good measurement of the model's overall ability to predict words
in the given corpus \cite{sundermeyer2015feedforward}. 
Perplexity for a given model is calculated
by exponentiating (base $e$) the mean
cross-entropy across all words in the test set. This is formally
expressed as follows:

\begin{equation}
\label{perplexity}
    P(L) = \exp\left(\frac{1}{N}\sum^{N}_{i=1} H(L,w_i)\right) ,
\end{equation}

where $N$ is the test data set size, $L$ is the language model, $w_i$
is the $i$th word in the test set, and $H(L, w_i)$ is the natural log
cross-entropy from $w_i$ to the prediction given by $L(w_i)$. A lower
perplexity represents a language model with better predictive
performance \cite{wang2016parallel}.

The cross-entropy is the opposite of summing the
product of the probability of that word appearing, i.e., $1$ for the
correct word and $0$ for all other incorrect words, and the
natural logarithm of the output value of LSTM's softmax layer.
The cross-entropy is defined as follows:

\begin{equation}
\label{cross-entropy}
    H(L,w) = - \sum_{i=1}^V p(w_i) \ln L(w_i) \, ,
\end{equation}

where $V$ is the vocabulary size and $p(w_i)$ is the probability of
$w_i$ being the correct word. Since the probability for incorrect
words is $0$ and the correct word is $1$, the sum can be reduced to
$-1$ times the natural log of the probability of the correct word
as given by the LSTM.
Thus, the cross-entropy is simply

\begin{equation}
\label{simple-cross-entropy}
    H(L,w) = - \ln L_w(w) \, .
\end{equation}

$L_w(w)$ represents the LSTM's softmax output specifically for the
word $w$. Additionally, mean word-level accuracy was calculated for
each language model considering the top $1$, $5$, and $10$ predictions
made by the model.

\section{Results}

\begin{table}[t]
    \caption{Perplexities ($P$) given by Equation \ref{perplexity}.
    Proportion of predictions which had the correct
    word in their top-$k$ predictions.
    ``ElasticSearch'' is written as ``ES'' and ``Spring
    Framework'' is written as ``SF''.}
    \label{topk-table}
    \vskip 0.15in
    \begin{center}
    \setlength{\tabcolsep}{\tabhorzspacing}
    {\renewcommand{\arraystretch}{\tabvertspacing}
    \begin{tabular}{lccccl}
    \hline
    Corpus & P & Top 1 & Top 5 & Top 10 & Language \\
    \hline
    PTB   & $85.288$ & $0.269$ & $0.470$ & $0.552$ & English \\
    JDK   & $21.808$ & $0.474$ & $0.652$ & $0.716$ & Java \\
    Guava & $18.678$ & $0.519$ & $0.696$ & $0.751$ & Java \\
    ES    & $11.397$ & $0.576$ & $0.739$ & $0.784$ & Java \\
    SF    & $11.318$ & $0.560$ & $0.722$ & $0.783$ & Java \\
    \hline
    \end{tabular}}
    \end{center}
    \vskip -0.1in
\end{table}

The perplexities achieved on the corpora by the LSTM are displayed in
Table \ref{topk-table}. The smallest perplexity for non-English data sets 
was measured for the Spring Framework, while the largest was for the JDK data. 
The table also indicates that all four Java data sets showed a drastic 
reduction in perplexity compared to the English data set. Nonetheless, the 
perplexity achieved on the English dataset is similar to that reported by 
Zaremba et al. \cite{Zaremba}. These results indicate the superiority of LSTMs on both
programming languages and a language as complex as the English language.

Table \ref{topk-table} shows the top-$k$ accuracies for each corpus.
Clearly, the results suggest that the proposed LSTM model is able to more
accurately model pre-processed Java source code than it can English. The table
also indicates that, for the English data set, the use of a large number of
predictors can dramatically increase the overall rate of predictors with the
correct next word; e.g., increasing from one to ten predictors at least doubled
the proportion of predictors. There is a similar effect over Java-based data
sets; however, in these data sets the predictors start at a higher proportion
than with English.

\section{Conclusion}

In this paper, we have presented a way of modeling a predictive strategy
over the Java programming language using an LSTM. Using datasets such as PTB,
JDK, Guava, ElasticSearch, and Spring Framework we have shown that
LSTMs are suitable in predicting the next syntactic statements of source
code based on preceding statements. Results indicate that indicate
that LSTMs can achieve lower perplexities and, hence, produce more accurate models
on the Java datasets than the English dataset.

The pre-processed Java code represents a very general and cursory
representation of the original code as it does not include anything such
as variable names or variable types. Future research along these lines
could account for information such as variable types, variable names, etc.
It would also be beneficial to compare the modeling of Java with other
programming languages or to train the model across multiple repositories
in one language.


\begin{thebibliography}{1}

\bibitem{Allamanis}
Miltiadis Allamanis and Charles Sutton.
\newblock Mining source code repositories at massive scale using language
  modeling.
\newblock In {\em Proceedings of the 10th Working Conference on Mining Software
  Repositories}, MSR '13, pages 207--216, Piscataway, NJ, USA, 2013. IEEE
  Press.

\bibitem{Nguyen}
Anh~Tuan Nguyen and Tien~N. Nguyen.
\newblock Graph-based statistical language model for code.
\newblock In {\em Proceedings of the 37th International Conference on Software
  Engineering - Volume 1}, ICSE '15, pages 858--868, Piscataway, NJ, USA, 2015.
  IEEE Press.

\bibitem{Asaduzzaman}
Muhammad Asaduzzaman, Chanchal~K. Roy, Kevin~A. Schneider, and Daqing Hou.
\newblock A simple, efficient, context-sensitive approach for code completion.
\newblock {\em Journal of Software: Evolution and Process}, 28(7):512--541,
  2016.
\newblock JSME-15-0030.R3.

\bibitem{Kim}
Dongsun Kim, Jaechang Nam, Jaewoo Song, and Sunghun Kim.
\newblock Automatic patch generation learned from human-written patches.
\newblock In {\em Proceedings of the 2013 International Conference on Software
  Engineering}, ICSE '13, pages 802--811, Piscataway, NJ, USA, 2013. IEEE
  Press.

\bibitem{Zaremba}
Wojciech Zaremba, Ilya Sutskever, and Oriol Vinyals.
\newblock Recurrent neural network regularization.
\newblock {\em CoRR}, abs/1409.2329, 2014.

\bibitem{Eclipse}
Eclipse Foundation.
\newblock Eclipse documentation on the {AST} class.
\newblock
  \url{http://help.eclipse.org/luna/index.jsp?topic=%2Forg.eclipse.jdt.doc.isv%2Freference%2Fapi%2Forg%2Feclipse%2Fjdt%2Fcore%2Fdom%2FAST.html},
  2016.
\newblock Accessed: 2016-8-18.

\bibitem{sundermeyer2015feedforward}
Martin Sundermeyer, Hermann Ney, and Ralf Schl{\"u}ter.
\newblock From feedforward to recurrent lstm neural networks for language
  modeling.
\newblock {\em IEEE/ACM Transactions on Audio, Speech and Language Processing
  (TASLP)}, 23(3):517--529, 2015.

\bibitem{wang2016parallel}
Minsi Wang, Li~Song, Xiaokang Yang, and Chuanfei Luo.
\newblock A parallel-fusion rnn-lstm architecture for image caption generation.
\newblock In {\em Image Processing (ICIP), 2016 IEEE International Conference
  on}, pages 4448--4452. IEEE, 2016.

\end{thebibliography}
\end{document}